\documentclass[fleqn,usenatbib]{mnras}

\usepackage{newtxtext,newtxmath}

\usepackage{hyperref}

\usepackage[T1]{fontenc}

\DeclareRobustCommand{\VAN}[3]{#2}
\let\VANthebibliography\thebibliography
\def\thebibliography{\DeclareRobustCommand{\VAN}[3]{##3}\VANthebibliography}

\usepackage{graphicx}	
\usepackage{amsmath}	
\usepackage{multirow}   
\usepackage{xcolor}

\title[UMa3/U1: star cluster or galaxy?]{Reevaluating UMa3/U1: star cluster or the smallest known galaxy?}

\author[S. Devlin et al.]{
Scot Devlin,$^{1}$\thanks{E-mail: s.devlin@uq.edu.au}
Holger Baumgardt,$^{1}$
Sarah M. Sweet$^{1, 2}$
\\
$^{1}$School of Mathematics and Physics, The University of Queensland, St. Lucia, QLD 4072, Australia\\
$^{2}$ARC Centre of Excellence for All Sky Astrophysics in 3 Dimensions (ASTRO 3D), Australia\\
}

\date{Accepted 2025 April 3. Received 2025 March 25; in original form 2024 December 13}

\pubyear{\the\year{}}

\begin{document}
\label{firstpage}
\pagerange{\pageref{firstpage}--\pageref{lastpage}}
\maketitle


\begin{abstract}
Ursa Major III/UNIONS 1 (UMa3/U1) is the faintest Milky Way satellite discovered to date, exhibiting a half-light radius of \(3 \pm 1 \, \text{pc}\) and an absolute V-band magnitude of \(+2.2 \pm 0.4\). Previous studies suggest UMa3/U1 is a dwarf galaxy, based on its large internal velocity dispersion and the improbability—indicated by dynamical cluster simulations—of its long-term survival if it were a dark-matter-free star cluster. In this paper, we model the evolution of UMa3/U1 as a star cluster using collisional N-body simulations that include a description of stellar evolution and the external tidal field of the Milky Way, with some simulations including primordial binaries. We find that UMa3/U1 has a substantial remaining lifetime of \(2.7 \pm 0.4 \, \text{Gyr}\), primarily due to the retention of compact stellar remnants within the cluster. This retention is facilitated by mass segregation and the preferential loss of low-mass stars. Furthermore, we demonstrate that the observed large velocity dispersion of UMa3/U1 can be successfully reproduced. These results support the possibility that UMa3/U1 is a self-gravitating star cluster. Our simulations reveal that modelling UMa3/U1 as a dark matter–free star cluster produces a markedly altered present-day mass function, driven by a strong depletion of low-mass stars. However, the degree of mass segregation among the visible stars is not statistically significant. We therefore recommend that future observations of UMa3/U1 and other very small Milky Way satellites focus on measuring their present-day mass functions to determine their nature.

\end{abstract}




\section{Introduction}
\label{Introduction}

Dwarf galaxies and star clusters are of significant cosmological interest as they are among the most numerous and smallest bound stellar systems in the Universe, providing key insights into galaxy formation and evolution. While they share similarities in size and stellar populations, they differ significantly in their dark matter content. Star clusters are largely devoid of dark matter and exhibit low mass-to-light ratios (\(1.4 < M/L_V < 2.5\)) \citep{Baumgardt_2020}.  Globular clusters, a type of star cluster typically found in the halos of galaxies, are believed to be ancient systems that have not undergone substantial chemical enrichment, making them valuable probes of the interstellar medium at the time of their formation \citep{Krumholz_2019}. These dense stellar systems can have central densities exceeding \(10^5 \, \mathrm{stars \, pc^{-3}}\), making them among the densest known stellar environments in the Universe \citep{Harris_1996}. In contrast, open clusters, which are younger and typically less massive systems located in the disks of galaxies, have much lower stellar densities, often ranging from \(1 \, \mathrm{star \, pc^{-3}}\) to \(100 \, \mathrm{stars \, pc^{-3}}\), and serve as tracers of recent star formation and the ongoing chemical evolution of their host galaxies \citep{Lada_2003}. More recently, low-density star clusters, resembling open clusters, have been discovered in the halo of the Milky Way \citep{Balbinot_2013, Kim_2015, Price-Whealan_2019}.

Dwarf galaxies are dark matter dominated; their dynamical masses far exceed their baryonic masses, as demonstrated by their exceptionally high mass-to-light ratios, typically around $M/L \sim 10^3 \left( \frac{M_{\odot}}{L_{\odot}} \right)$ \citep{Simon_2019}. Dwarf galaxies are excellent small-scale dark matter probes because they contain just enough visible matter to trace the distribution of invisible dark matter, without having so much visible matter that it significantly contributes to the total galaxy mass. The ability to trace the dark-matter distribution without significant baryonic mass contribution is especially true for the smallest dwarf galaxies, known as ultra-faint dwarf galaxies (UFDs). UFDs provide crucial information on the minimum dark-matter halo mass required for galaxy formation \citep{Bode2001}. Moreover, the absolute number of UFDs places constraints on star formation theories \citep{Munshi_2019} and on the mass of the dark matter particle \citep{Simon_2019}. In addition, the position of UFDs within the Milky Way halo helps to constrain the “Satellite Plane Problem” \citep{Kroupa_2005, Pawlowski_2012, Pawlowski_2013}. Because UFDs are less influenced by baryonic processes than larger galaxies, their density profiles offer robust tests of the cusp-core problem, thereby probing the validity of various cosmological models \citep{almeida_2024}.

Distinguishing between star clusters and UFDs is an important task, especially given the rapid increase in the discovery rate of faint Milky Way satellites since the early 2000s, driven by wide-field photometric surveys \citep{Simon_2019}. For stellar systems in the Milky Way halo brighter than \( M_V \approx -5 \), a clear size difference generally exists between star clusters and dwarf galaxies. Bright star clusters have half-light radii \( r_h < 20 \, \mathrm{pc} \), while dwarf galaxies typically have \( r_h > 100 \, \mathrm{pc} \) \citep{Simon_2019}. For slightly fainter stellar systems, the magnitude-size plane can still be useful, as star clusters are known to be more compact than dwarf galaxies, which allows for a rough classification of stellar systems based on their surface or 3D density; see figure 2 of \citet{Simon_2019}, figure 6 of \citet{Baumgardt_22}, and figure 9 of \citet{Smith_2024}. However, it has become increasingly challenging to distinguish between star clusters and dwarf galaxies at the fainter end of the distribution based solely on their radii and luminosities, prompting the need for additional classification criteria.

In addition to having a dynamical mass that far exceeds their baryonic mass, UFDs are distinguished from star clusters by exhibiting a non-zero spread in metallicities \citep{Willman_2011}. The metallicity spread seen in UFDs arises because the dark matter halo creates a deeper gravitational potential well, which raises the escape velocity for gas in these systems. The higher escape velocity allows UFDs to more effectively retain chemically enriched gas (e.g., supernova ejecta) released during stellar explosions. Enhanced gas retention, in turn, supports extended star formation periods and sustained chemical self-enrichment.

The recently identified Ursa Major III/UNIONS 1 system (UMa3/U1 or U1 if referring to the system as a star cluster and UMa3 if referring to the satellite as an ultra-faint dwarf galaxy), discovered through the deep, wide-field Ultraviolet Near Infrared Optical Northern Survey (UNIONS), is the least luminous known satellite of the Milky Way. It has an absolute V-band magnitude of \( M_V = 2.2^{+0.3}_{-0.4} \) and is located at a heliocentric distance of \( 10 \pm 1 \) kpc \citep{Smith_2024}. It is 1.5 magnitudes fainter than the next faintest Milky Way satellites discovered so far, the ultrafaint dwarfs Kim 3 (\( M_V = +0.7 \) mag; \citealt{Kim2016}) and DELVE 5 (\( M_V = +0.4 \) mag; \citealt{Cerny2023}). The small size of UMa3/U1, with a half-light radius of \( r_h = 3 \pm 1 \ \text{pc} \), is well within the size range of known star clusters and is an order of magnitude smaller than the smallest confirmed ultra-faint dwarf galaxy, Willman 1, which has a half-light radius of approximately, \( r_h \approx 20 \ \text{pc} \) \citep{Willman_2011}. However, \citet{Smith_2024} argue that the high line of sight velocity dispersion, \( \sigma_{los} = 3.7^{+1.0}_{-1.4} \, \text{km} \, \text{s}^{-1} \) they observe suggests that UMa3/U1 is a dark matter-dominated galaxy. Similarly, \citep{Errani_2024} analytically estimate that if UMa3/U1 were a self-gravitating star cluster devoid of dark matter, its velocity dispersion should be only $\sigma_{los} = 0.049^{+0.014}_{-0.011}  \text{km} \, \text{s}^{-1}$, significantly lower than the observed value. \citet{Errani_2024} also argue that, if UMa3/U1 were a star cluster, its low density would cause it to quickly disintegrate due to tidal interactions with the Milky Way's gravitational potential. Using the collisionless particle mesh code, \textsc{Superbox} \citep{Fellhauer_2000_SUPERBOX}, they demonstrate that UMa3/U1 could not survive for much longer than 0.4 Gyr as a dark matter free star cluster. 0.4 Gyr is short compared to its proposed age of \(\approx 12\) Gyr \citep{Smith_2024}, observing UMa3/U1 just before its final demise is unlikely. Thus, they conclude that UMa3/U1 is more likely a UFD protected from tidal stripping by its own dark matter halo. 

The total mass of a dispersion-supported star system can be reliably constrained by its velocity dispersion only if three key assumptions are met: the system is in dynamical equilibrium, contamination by binary stars is minimal, and contamination by foreground and background stars is negligible \citep{Simon_2019}. While foreground, background, and tidally unbound stars can be largely filtered out using isochrone methods, accurately identifying binary stars requires multiepoch spectroscopic surveys with an appropriate cadence. \citet{McConnachie_2010} find that unidentified binaries do not significantly increase the velocity dispersion of globular clusters beyond about $\sigma_{los} = 4.5 \text{km} \, \text{s}^{-1}$. Hence, for larger UFDs and most globular clusters, the impact of stellar binaries on overall velocity dispersion is minimal \citep{Simon_2019}. However, with only 11 kinematically confirmed member stars \citep{Smith_2024}, UMa3/U1 is much more susceptible to artificial inflation of its velocity dispersion due to small number statistics. \citet{Smith_2024} note this issue and show that the removal of the star with the largest outlying velocity would result in a velocity dispersion of $ \sigma_{los} = 1.9^{+1.4}_{-1.1} \, \text{km} \, \text{s}^{-1}$, still large enough to constitute a dynamical mass three orders of magnitude larger than the stellar mass, when using the \citet{Wolfe_2010} dynamical half mass formula. Removing the second largest outlier as well results in a formally unresolved velocity dispersion, consistent with what would be expected if UMa3/U1 is a self-gravitating star cluster.    

If UMa3/U1 is a star cluster an estimate of the expected fraction of binaries can be obtained by extrapolating trends seen for clusters with observable binary systems: Although binaries in globular clusters are challenging to observe directly due to crowding and the faintness of stars, cluster binary fractions have been estimated by identifying binaries above the main sequence in colour-magnitude diagrams. \citet{Milone_2012} reported an average binary fraction of $f_{\text{b}} \approx 10\% $ in globular clusters, with higher fractions $f_{\text{b}} > 40\%$ in lower mass systems like Palomar 1 and E 3. \citet{Anjana_2024} noted an anticorrelation between binary fraction and cluster mass, suggesting that smaller clusters such as U1 could exhibit higher binary fractions. Lower-density clusters reduce binary disrupting interactions, also supporting higher binary fractions \citep{Sollima_2007}. Even though binary fractions decrease with metallicity \citep{El-Badry_2019, Offner_2023} --- as dense, low-metallicity environments disrupt wider binaries \citep{Hwang_2021} --- the small size and low density of UMa3/U1 (much smaller than Palomar 1 and E 3) suggest that it likely has a high present-day binary fraction.

Confirming UMa3/U1 as a dark matter-dominated dwarf galaxy would be of significant cosmological interest. Its proximity, extreme faintness, and unmatched compactness make it a prime candidate for hosting the densest dark matter halo. Therefore UMa3/U1 is an excellent target for probing dark matter, such as through the indirect detection of gamma rays from dark matter annihilation \citep{Andrade_2023}.

In this paper, we present collisional modelling of the tidal evolution of UMa3/U1 as a star cluster using \textsc{Nbody7}, in order to investigate the impact that tidal forces and stellar evolution have on UMa3/U1's velocity dispersion and time to dissolution. We also look at two alternate ways of differentiating star clusters from UFDs, and apply them to our U1 models: stellar mass segregation \citep{Baumgardt_22}, and the present-day mass function. 

This work is organised as follows: Section \ref{methods} describes how we model UMa3/U1 and explains how the model velocity dispersions are calculated. Section \ref{time_dis_sect} discusses the results for U1's time to dissolution, and compares them to previous published results. Section~\ref{density} discusses how dynamical processes contribute to an increase in our model cluster densities, thereby extending their remaining lifetimes. Section \ref{vel_dispers} compares the velocity dispersion results to analytical velocity dispersion values and explains the necessity of primordial binaries. Section \ref{mass_seg_section} presents the mass segregation of the model cluster, and discusses its usefulness in classifying UMa3/U1. Section \ref{mass_funct_sect} describes the method and rationale for modeling the present-day mass function of UMa3/U1, and discusses its usefulness as a parameter for classifying UMa3/U1, as well as other faint Milky Way satellites. A summary of our results, as well as a discussion of the limitations of our methods is provided in Section~\ref{summarry_sect}.

\section{Method}
\label{methods}
\subsection{N-Body simulations}
\label{N-body}
The simulations were carried out with the collisional Aarseth N-body code \textsc{Nbody7} \citep{Nitadori_and_Aarseth_2012}. The code uses a Hermite scheme with individual time-steps for the integration and handles close encounters between stars using KS \citep{Kustaanheimo_1965} and chain regularizations \citep{Mikkola_1989, Mikkola_1993}. Our simulations model stellar evolution using the fitting formulas of \citet{Hurley_2000}, which provide stellar lifetimes, luminosities, and radii as functions of initial mass and metallicity across all phases of stellar evolution, from the zero-age main sequence through to the compact stellar remnant stages.

To understand how long UMa3/U1 will survive if it is a star cluster, it is necessary to simulate the effects of tidal stripping and stellar evolution on U1 from the cluster's birth. Although \textsc{Nbody7} can precisely calculate the effects of tidal stripping by tracking the position of the cluster relative to the Milky Way over time, the initial position and velocity of the cluster relative to the Milky Way potential are required to begin the simulation. We calculate the initial position and velocity by taking the six phase-space coordinates describing the present-day position and motion of UMa3/U1, as provided in table 3 of \citet{Smith_2024}, and used the \textsc{Galpy} program \citep{Bovy_2015} to integrate a massless test particle backward in time for a duration of 12 Gyr under the influence of the \citet{Bovy_2015} Milky Way potential. To convert from equatorial to Galactic Cartesian coordinates, we used the solar velocity relative to the Local Standard of Rest (LSR) from \citet{Schonrich_2010}, assuming a solar distance to the Galactic center of 8 kpc, a solar height above the Galactic plane of 25 pc, and a circular velocity of the LSR around the Galactic center of $220\,\mathrm{km\,s^{-1}}$. This provides the starting position and velocity for the N-body simulation: $X = 17.404 \, \text{kpc}$, $Y = -19.215 \, \text{kpc}$, $Z = -19.285 \, \text{kpc}$, and velocities $U = 20.418\,\mathrm{km\,s^{-1}}$, $V = 141.333\,\mathrm{km\,s^{-1}}$, and $W = 10.903\,\mathrm{km\,s^{-1}}$. We also add the Milky Way potential from \citet{Bovy_2015} to \textsc{Nbody7} to allow a forward
integration of the full cluster model on the same orbit as the
backward integration was performed.

\subsection{Replicating UNIONS and \textit{Hubble Space Telescope (HST)} observations}
\label{sec:Unions_range}
To facilitate a direct comparison between our simulations and both current and future observations of UMa3/U1, we design two filters that emulate the capabilities of the UNIONS survey and a deeper photometric survey, such as one that could be executed with the (\textit{HST}). We refer to the stars remaining after applying each respective filter as belonging to the "UNIONS range" and the "\textit{HST} range."

\citet{Smith_2024} use $i$-band data from the Pan-STARRS survey. Pan-STARRS has a saturation limit of \( i = 17.5 \) mag and achieves a median \( 5\sigma \) point source depth of \( i = 24.0 \) mag, from which \citet{Smith_2024} adopt a completeness limit of \( i = 23.5 \) mag. We estimate which stars \citet{Smith_2024} would have been able to observe in our model U1 clusters by removing: (a) stars fainter than the adopted completeness limit of apparent $i$-band magnitude, \(i = 23.5 \, \text{mag}\) and stars brighter than the saturation limit of \(i = 17.5 \, \text{mag}\) \citep{Smith_2024}; (b) compact stellar remnants; and (c) stars that are not located within an ellipse centered on the density center, with a semi-major axis of four times the half-light radius (\(4 \times r_h = 12 \, \text{pc}\)) and an ellipticity of \(\epsilon = 0.5\) (Table 2 of \citealt{Smith_2024}). We refer to these selection criteria as the `UNIONS range.'

We apply criterion (a) by excluding stars with masses higher than \(0.79\,M_\odot\) and lower than \(0.31\,M_\odot\), the mass equivalents to 17.5 and 23.5 in apparent $i$-band magnitude, as measured by the Pan-STARRS survey. To obtain the conversion from apparent $i$-band magnitudes to equivalent masses, we use the Padova and Trieste Stellar Evolution Code (PARSEC) 1.2 track isochrones \citep{Bressan_2012}, utilising the \href{http://stev.oapd.inaf.it/cgi-bin/cmd}{CMD 3.7} tool to find the best-match isochrone to the following parameters:
\begin{enumerate}
    \item An extinction of \(A_v=0.0584\) from a colour excess \(E(B-V)=0.0213\), obtained using the extinction tool, \href{http://www.galextin.org/}{GALExtin} \citep{Amôres_21}, employing the \citet{Schlegel_1998} extinction map for the location of UMa3/U1 from \citet{Smith_2024};
    \item A metallicity of \(Z = 0.0001\) (rounded up from 0.0000959 because PARSEC isochrones are only available for metallicities larger than \(Z = 0.0001\)), equivalent to the \([Fe/H] = -2.2\) of UMa3/U1 found by \citet{Smith_2024} for solar metallicity of \(Z = 0.0152\), and assuming that the [Fe/H] fraction scales with overall metallicity [M/H];
    \item Age of 12 Gyr from \citet{Smith_2024}.
\end{enumerate}

To create the \textit{HST} filter, we exclude all compact stellar remnants—white dwarfs, neutron stars, and black holes—as they are too faint to be detected in deep photometric observations possible with telescopes like the \textit{(HST)} or the \textit{James Webb Space Telescope (JWST)}

\subsection{Setting up the UMa3/U1 progenitor}
\label{progen_model}
The models used for the progenitor of U1 are constructed as spherically symmetric \citet{King_1962} models assuming an isotropic velocity distribution and no primordial mass segregation. They all have a concentration parameter of \( c = 1 \) and a metallicity of \( Z = 0.0001 \) (see Section \ref{sec:Unions_range}). All simulated progenitors begin with a physical half-light radius of, $r_h=3~pc$, as trial simulations starting at $r_h=3~pc$ consistently produce a final half-light radius of approximately $3~pc$ - matching the observed half-light radius of UMa3 / U1 \citep{Smith_2024}. We create the progenitors using the program described in \citet{Hilker_2007} which sets up equilibrium models of stellar systems by calculating their distribution function.  We use the initial mass function (IMF) defined by \citet{Baumgardt_2023}, a three-stage power law with a near-Salpeter-like \citep{Salpeter_1955} slope above 1.0 \(M_{\odot}\), a shallow slope below 0.4 \(M_{\odot}\), and an intermediate slope between 0.4 and 1.0 \(M_{\odot}\):
\begin{equation}
\label{IMF}
\xi(m) \, dm \sim
\begin{cases} 
m^{-0.3} \, dm, & \text{for } m < 0.4 \, M_{\odot} \\
m^{-1.65} \, dm, & \text{for } 0.4 \, M_{\odot} \leq m < 1.0 \, M_{\odot} \\
m^{-2.3} \, dm, & \text{for } m \geq 1.0 \, M_{\odot}
\end{cases}
\end{equation}

We conduct two sets of simulations with assumed 0\% and 10\% black hole (BH) and neutron star (NS) retention fractions. In this context, the retention fraction refers to the cluster’s ability to keep a BH or NS gravitationally bound after it receives an asymmetric kick from its progenitor supernova. Low retention fractions are chosen because the relatively small mass of our starting cluster implies an escape velocity of only 4 to 5 $\text{km}\,\text{s}^{-1}$—a value significantly lower than the predicted neutron star natal kick velocities for most compact remnant masses and most asymmetric supernova mechanisms \citep[][figure 11]{Banerjee_2020}. We model natal kicks for BHs and NSs by adding a strong enough velocity kick to every newly formed BH and NS so that the stars become unbound and leave the cluster. BHs and NSs that are retained do not receive a velocity kick upon formation. Whether or not a star is removed is randomly decided upon formation depending on the chosen retention fraction.

To select the appropriate mass for the U1 progenitor, we perform a series of trial simulations. Each simulation starts from the initial position and velocity described in Section \ref{N-body}. In these simulations, the number of stars in the progenitor cluster, \(N\), is varied between 1,000 and 30,000, while \(r_h\), \(c\), and \(Z\) remain fixed at the values mentioned above. The progenitor size \(N\) for the simulations to be analysed is determined by selecting the trial simulation whose remaining star count at an age of 12~Gyr most closely matches the 21 photometric member stars observed for UMa3/U1 down to a Pan-STARRS \(i\)-band magnitude of 23.5~mag by \citet{Smith_2024} within the UNIONS range (Section~\ref{sec:Unions_range}). We find that the progenitor of UMa3/U1 should typically have \(N=6000\) stars if all BHs and NSs are given natal kick velocities above the cluster escape velocity and \(N=7200\) if 10\% of the BHs and NSs are retained.

We then run multiple simulations with either a 0\% or 10\% retention rate of NSs and BHs, until we obtain ten simulations for each case that have 21 stars in the UNIONS range. We allow a time window between 10 Gyr and 14 Gyr to reach 21 stars for the simulations to reflect the uncertainty in the cluster age. The results of simulations that meet the time window criteria are given in Table 1. The first 10 simulations with a 0\% retention fraction reached 21 stars in the UNIONS range in the target time window, many coming very close to the target age of 12 Gyr. The clusters with a 10\% retention fraction exhibit greater variability in their lifetimes and the time to reach 21 stars, because of the more unpredictable evolution of clusters with a population of massive, stellar-mass black holes. In these simulations, it is not uncommon for the ejection of a single black hole binary to lead to the rapid dissolution of the cluster, with the timing of such events being stochastic. As a result we conduct 19 simulations with a 10\% retention rate to obtain 10 that fall within the 10 - 14 Gyr time window.

\subsection{Bound Star Method: Tidal Radius and Density Centre Calculations}
\label{bound star method}
To understand how many stars remain bound in the cluster at a particular time, the density centre is determined using the method of \citet{Casertano_1985}, calculating the local density around each star by taking the inverse cube of the distance to the 7th nearest neighbour. To calculate the tidal radius we use the formula for a cluster on a circular orbit within a spherically symmetric isothermal external potential \citep{King_1962, Baumgardt_2003}:
\begin{equation}
r_t = \left( \frac{G M_c}{2 V_G^2} \right)^{1/3} R_G^{2/3}
\label{circular tidal radius}
\end{equation}
where \( M_c \) is the mass of the cluster, \( V_G \) is the circular velocity of the Galaxy, and \( R_G \) is the distance of the cluster from the Galactic Centre. For the eccentric orbit of UMa3/U1 we set \( V_G \) to be the instantaneous orbital velocity and \( R_G \) to be the instantaneous distance from UMa3/U1 to the centre of the Milky Way.
In some simulations, the clusters survive with fewer than 8 stars for $>100$ Myr. In these cases the \citet{Casertano_1985} method is adjusted for the time steps where fewer than 8 stars are bound to take the inverse cube of the distance to the 7th, 6th, or 5th nearest neighbour of each bound star, thereby allowing the density centre to be computed for smaller cluster sizes.

At the start of the simulations all stars are assumed to be bound. In subsequent time steps, we initially calculate the density centre and cluster mass, \( M_c \), from the position and mass of the bound stars of the previous time step. \( R_G \) is then calculated from the new density centre, and a new tidal radius is calculated. The new tidal radius is then used to determine a new set of bound stars.
We then update our density centre, \( M_c \) and \( R_G \) using only the new set of bound stars, and repeat this procedure until the calculated cluster parameters converge.

\begin{figure*}
    \centering
    \begin{minipage}{1.0\textwidth}
        \centering
        \includegraphics[width=\textwidth]{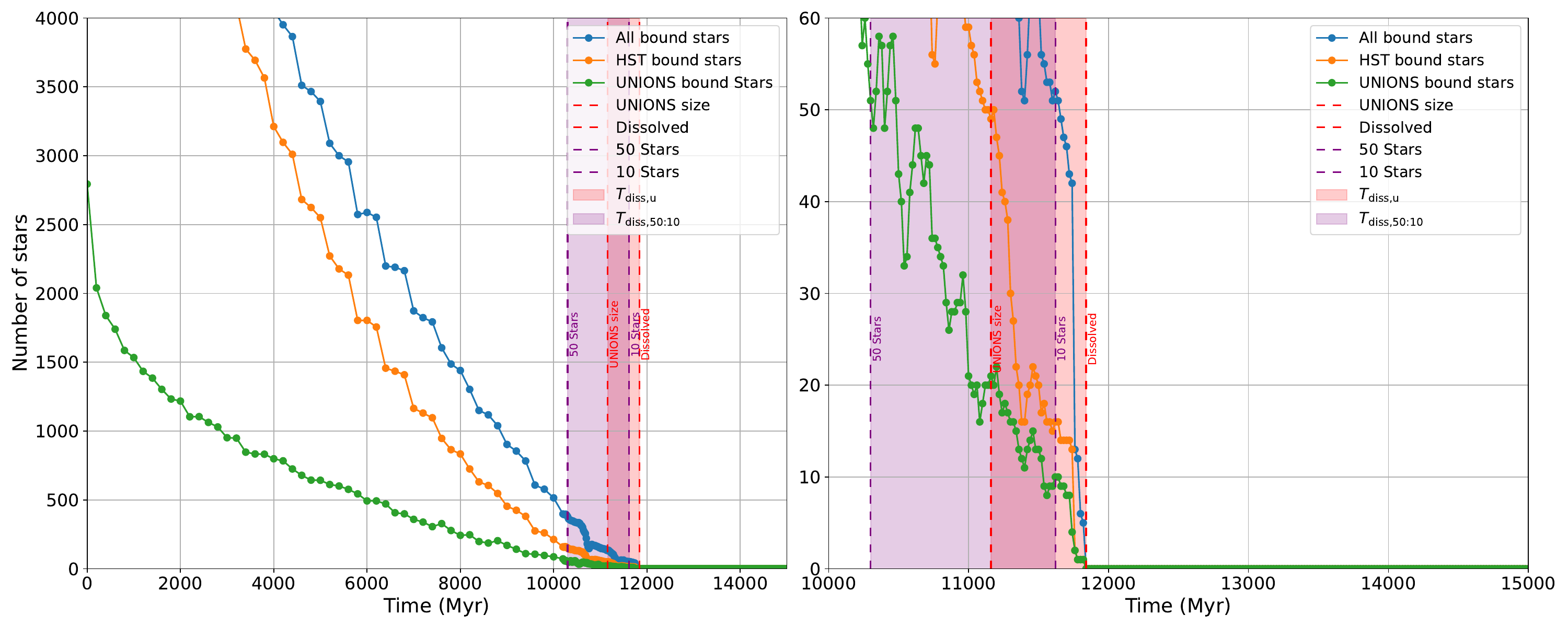}
    \end{minipage}
    \vfill
    \begin{minipage}{1.0\textwidth}
        \centering
        \includegraphics[width=\textwidth]{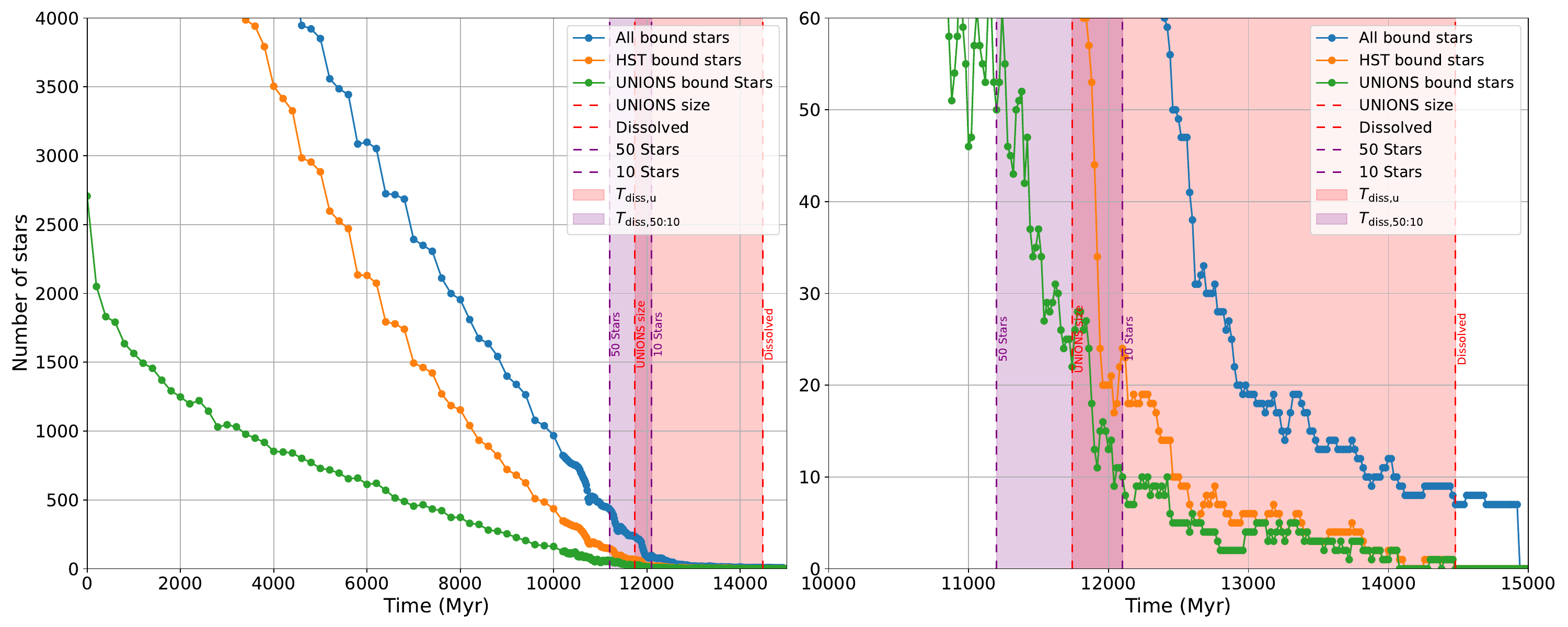}
    \end{minipage}
    \caption{Number of bound stars in simulation 12 (top panels) and simulation 17 (bottom panels). The green lines represent the number of bright stars visible to a UNIONS-like survey, the orange line represents the number of bound main-sequence and giant stars, and the blue line represents the total number of bound stars as a function of time. The red vertical dashed lines bounding the red shaded area mark the remaining lifetime of the U1 simulated cluster, from its current size to its final dissolution in the bright stars (visible to a UNIONS-like survey). The purple vertical dashed lines bounding the purple shaded area marks the time span in which the U1 simulated cluster would have 50 to 10 stars visible to the UNIONS survey. Simulation 17 depicts an evolution typical of most simulations, characterized by gradual changes in the cluster's structure. In contrast, simulation 12 exhibits an unusually abrupt disruption, driven by the ejection of a massive black hole binary at 11740 Myr.}

    \label{fig:time_evo}
\end{figure*}

We calculate the remaining UNIONS range lifetime, $T_{diss,U}$ of each cluster by taking the time when the cluster most closely resembles UMa3/U1 (we call this UNIONS time) up to the time when the cluster is finally dissolved. Both UNIONS time and cluster dissolution time are represented by vertical red dashed lines in Figure \ref{fig:time_evo}, a plot of cluster size vs time. The cluster is classified as dissolved when too few stars are in the cluster to calculate the density centre (8 stars or less for the unadjusted density centre calculations). The cluster is considered to most closely resemble UMa3/U1 when 21 stars are visible in the UNIONS range. Due to the coarseness of our simulation time sampling (initially every 200 Myr, then every 20 Myr after 10 Gyr, except for simulations 21 and 22, which maintain 200 Myr intervals throughout due to computational constraints), clusters often reached 21 stars between time steps. In these cases, we select the timestep at which the cluster membership is closest to 21 stars within the UNIONS range. In all cases, the cluster is considered to be at 'UNIONS size' once it reaches UNIONS time. In some simulations, clusters exhibit 21 bound stars in UNIONS range stars at multiple time steps. When this occurs, we introduce a secondary criterion, the half light radius: out of the multiple time steps with 21 stars we pick the time step when the cluster has a half light radius closest to UMa3/U1's 3pc half light radius \citep{Smith_2024}.

\subsection{Velocity dispersion}
\label{velocity dispersion method}

Simulations 1 to 20 (see Table~1) are initiated without primordial binaries (\(f_{b,0} = 0\%\)). We chose not to include primordial binaries in these simulations because their presence significantly increases the computational expense of N-body calculations \citep{Trenti_2006}. Although primordial binaries can affect the internal dynamics of a cluster, our primary objective, to estimate the remaining lifetime of the U1 star cluster (a timescale primarily governed by the strength of the tidal field and the orbital parameters of the cluster \citep{Baumgardt_2003}) is relatively insensitive to the initial binary fraction \citep{Vesperini_1997}. Moreover, any binaries or higher multiples that form dynamically during integrations are assumed to neither collide nor exchange mass. We further assume that the stellar evolution of the individual components remain unaffected by their companions. These assumptions are justified by the large semi-major axes of dynamically forming binaries. We find that the number of dynamically formed binaries in the \(f_{b,0} = 0\%\) simulations is very low (see the \(f_{b,\text{all}}\) column in Table~\ref{22_sims}), and insufficient to increase the cluster's velocity dispersion (see Section~\ref{vel_dispers} for further discussion).

To examine the impact of primordial binaries on the velocity dispersion, we conduct four additional simulations, Simulations 21 - 24 in Table \ref{22_sims}, with primordial binary fractions of 50\%.  These simulations are set up as described in Section~\ref{progen_model} using the \citet{Baumgardt_2023} IMF, which produces stars with an average initial mass close to one $M_{\odot}$. The binaries are initialised with a flat distribution in the semi-major axis and with masses paired at random. We adopt \( f_{\text{b,0}} = 50\% \) because nearby field star surveys, primarily sampling stars from the relatively young Galactic disk and thus reflective of primordial star cluster conditions, show that for primary stars of approximately one $M_\odot$, the observed multiplicity fraction is roughly \(50\%\) (see Fig.~1 in \citet{Offner_2023}).

We calculate the component velocities of all stars in the UNIONS range relative to the cluster centre of mass, \( V_x \), \( V_y \), and \( V_z \), and then merge all three components into a single list, taking the standard deviation of the merged list as the velocity dispersion of the cluster. This method improves statistical reliability by increasing the number of data points. Since globular clusters are generally expected to be nearly isotropic in their velocity distributions \citep{King_1966, Meylan_1996}, combining all three velocity components provides a robust measure of overall velocity dispersion. By individually evaluating the velocity dispersion components \(V_x\), \(V_y\), and \(V_z\), we find that their deviations from the overall mean are well within the error bars of the UMa3/U1 velocity dispersion \citep{Smith_2024}. Consequently, any anisotropy in the simulations does not significantly affect the velocity dispersion of our merged list. Due to the challenge of resolving closely separated binaries at the distance of UMa3/U1 (10 kpc), we treat all binaries as single stars, assigning them a velocity equal to the luminosity-weighted average velocity of their two components:
\begin{equation}
v_{\text{lum,j}} = \frac{L_{A,j} v_{A,j} + L_{B,j} v_{B,j}}{L_{A,j} + L_{B,j}}
\end{equation}
where \( L_{A,j} \) and \( L_{B,j} \) are the bolometric luminosities of the individual component stars in the binary pair, such that their sum gives the total bolometric luminosity of the system. Similarly, \( v_{A,j} \) and \( v_{B,j} \) denote the velocities of the two stars comprising the binary.
In the presence of binaries, the velocity dispersion of the cluster is calculated as in \citet[][Method 3]{Rastello_2020}:
\begin{equation}
\sigma_{lum} = \sqrt{\frac{1}{N_s + N_b} \left( \sum_{i=1}^{N_s} v_i^2 + \sum_{j=1}^{N_b} v_{\text{lum,j}}^2 \right)}
\end{equation}
where \( N_s \) is the number of single stars and \( N_b \) is the number of binaries.

\section{Results and Analysis}

In this section, we present our results from the simulations aimed at determining the nature of UMa3/U1: In Section \ref{time_dis_sect}, we calculate the remaining lifetime of the clusters in our simulations. In Section \ref{density}, we compare the density of the clusters in our simulations with the density of the Milky Way inside the orbit of U1. In Section \ref{vel_dispers}, we compare the cluster velocity dispersion with analytical velocity dispersion values, and investigate the effect of primordial binaries on the velocity dispersion. In Section \ref{mass_seg_section}, we examine mass segregation as a potential classification tool. Finally in Section \ref{mass_funct_sect}, we explore the mass function as a classification tool for ultra-faint dwarfs (UFDs) and star clusters.

\begin{table*}
\centering
\caption{Summary of our simulations: \(s\) is the simulation number. N is the number of initial stars. RR is the adopted retention rate for black holes and neutron stars. \(f_{b0}\) is the fraction of primordial binaries. \(f_{b,all}\) is the fraction of bound binaries at UNIONS time. \(f_{b,U}\) is the fraction of bound binaries visible in the UNIONS survey. \(T_{\text{diss,U}}\) is the remaining lifetime of the cluster; from 21 to 0 stars visible to the UNIONS survey. \(T_{\text{diss,U,50:10}}\) is the remaining lifetime of the cluster; from 50 to 10 stars visible to the UNIONS survey. \(\sigma_{\text{lum,U}}\) is the velocity dispersion of UNIONS visible stars, using the luminosity weighted average velocity for binaries. \(\sigma_{\text{sing,U}}\) represents the velocity dispersion of all single stars observable by UNIONS. \(p_{\text{HST}}\) and \(p_{\text{U}}\) are the Kolmogorov–Smirnov (KS) test \(p\)-values obtained by comparing the UMa3 UFD present-day model mass function with the present-day mass function of the simulated clusters. Specifically, \(p_{\text{HST}}\) is computed when both mass functions are restricted to the \textit{HST} range (i.e. all stars except compact remnants), while \(p_{\text{U}}\) is determined when the comparison is limited to the UNIONS range. \(\rho_{all}\) is the density of all bound stars within the half-light radius. \(\rho_{HST}\) is the density of all stars except compact remnant stars within the half-light radius. \(f_{cr}\) is the fraction of compact remnants in the cluster at UNIONS time. There is no \(\sigma_{\text{sing,U}}\) value for simulation 23 because $100\%$ of the UNIONS range stars are in binaries.}

\label{22_sims}
\resizebox{\textwidth}{!}{
\begin{tabular}{|c@{\hspace{0.4cm}}c@{\hspace{0.4cm}}c@{\hspace{0.4cm}}c@{\hspace{0.4cm}}c@{\hspace{0.4cm}}c@{\hspace{0.4cm}}c@{\hspace{0.4cm}}c@{\hspace{0.4cm}}c@{\hspace{0.4cm}}c@{\hspace{0.4cm}}c@{\hspace{0.4cm}}c@{\hspace{0.4cm}}c@{\hspace{0.4cm}}c@{\hspace{0.4cm}}c@{\hspace{0.4cm}}|}
\hline
\multirow{2}{*}{$\textbf{s}$}  & \multirow{2}{*}{$\textbf{N}$} & \textbf{RR} & \textbf{\boldmath{$f_{b0}$}} & \textbf{\boldmath{$f_{b,\text{all}}$}} & \textbf{\boldmath{$f_{b,\text{U}}$}} & \textbf{$T_{\text{diss,U}}$} & \textbf{$T_{\text{diss,U,50:10}}$} & \textbf{$\sigma_{\text{lum,U}}$} & \textbf{$\sigma_{\text{sing,U}}$} & \multirow{2}{*}{$p_{\text{\scriptsize HST}}$} &  \multirow{2}{*}{$p_{\text{\scriptsize U}}$} & \textbf{\boldmath{$\rho_{\text{all}}$}} & \textbf{\boldmath{$\rho_{\text{\scriptsize HST}}$}} & \textbf{\boldmath{$f_{\text{cr}}$}} \\
 &  &\textbf{(\%)} &\textbf{(\%)} & \textbf{(\%)} & \textbf{(\%)} & \textbf{(Myr)} & \textbf{(Myr)} & \textbf{(Myr)} & \textbf{\boldmath($\mathrm{km\,s^{-1}}$)} &  &  & \textbf{($M_{\odot}$kpc$^{-3}$)} & \textbf{($M_{\odot}$kpc$^{-3}$)} & \textbf{(\%)} \\
\hline
1 & 6000 & 0 & 0 & 2 & 0 & 2080 & 1900 & 0.15 & 0.15 & $8.10 \times 10^{-13}$ & $1.24 \times 10^{-4}$ & $4.00 \times 10^{8}$ & $4.01 \times 10^{7}$ & 75 \\
2 & 6000 & 0 & 0 & 2 & 0 & 2220 & 980 & 0.16 & 0.16 & $5.41 \times 10^{-23}$ & $2.56 \times 10^{-1}$ & $9.02 \times 10^{8}$ & $1.07 \times 10^{8}$ & 72 \\
3 & 6000 & 0 & 0 & 1 & 0 & 2361 & 2000 & 0.13 & 0.13 & $2.17 \times 10^{-15}$ & $1.39 \times 10^{-3}$ & $3.68 \times 10^{8}$ & $4.65 \times 10^{7}$ & 76 \\
4 & 6000 & 0 & 0 & 1 & 0 & 2800 & 2200 & 0.17 & 0.17 & $7.79 \times 10^{-16}$ & $4.80 \times 10^{-5}$ & $4.32 \times 10^{8}$ & $5.38 \times 10^{7}$ & 78 \\
5 & 6000 & 0 & 0 & 1 & 0 & 2220 & 1760 & 0.16 & 0.16 & $2.66 \times 10^{-18}$ & $3.66 \times 10^{-5}$ & $4.68 \times 10^{8}$ & $6.56 \times 10^{7}$ & 79 \\
6 & 6000 & 0 & 0 & 0 & 0 & 2160 & 1880 & 0.15 & 0.15 & $1.76 \times 10^{-12}$ & $8.56 \times 10^{-5}$ & $5.83 \times 10^{8}$ & $6.11 \times 10^{7}$ & 78 \\
7 & 6000 & 0 & 0 & 3 & 5 & 2480 & 1680 & 0.15 & 0.14 & $2.73 \times 10^{-18}$ & $5.40 \times 10^{-4}$ & $3.29 \times 10^{8}$ & $5.58 \times 10^{7}$ & 79 \\
8 & 6000 & 0 & 0 & 2 & 0 & 2820 & 1320 & 0.16 & 0.16 & $1.73 \times 10^{-17}$ & $5.02 \times 10^{-4}$ & $3.35 \times 10^{8}$ & $6.44 \times 10^{7}$ & 78 \\
9 & 6000 & 0 & 0 & 2 & 0 & 2900 & 1940 & 0.16 & 0.16 & $3.63 \times 10^{-8}$ & $4.85 \times 10^{-3}$ & $2.20 \times 10^{8}$ & $3.21 \times 10^{7}$ & 82 \\
10 & 6000 & 0 & 0 & 1 & 5 & 1980 & 1780 & 0.16 & 0.16 & $2.37 \times 10^{-13}$ & $2.73 \times 10^{-3}$ & $5.09 \times 10^{8}$ & $6.72 \times 10^{7}$ & 80 \\
11 & 7200 & 10 & 0 & 1 & 0 & 2380 & 1720 & 0.15 & 0.15 & $5.41 \times 10^{-19}$ & $7.73 \times 10^{-3}$ & $2.96 \times 10^{8}$ & $2.72 \times 10^{7}$ & 77 \\
12 & 7200 & 10 & 0 & 2 & 0 & 680 & 1320 & 0.51 & 0.51 & $3.88 \times 10^{-15}$ & $8.94 \times 10^{-6}$ & $5.47 \times 10^{8}$ & $4.36 \times 10^{7}$ & 63 \\
13 & 7200 & 10 & 0 & 1 & 0 & 120 & 180 & 0.15 & 0.15 & $4.85 \times 10^{-21}$ & $1.15 \times 10^{-4}$ & $3.60 \times 10^{8}$ & $4.35 \times 10^{7}$ & 59 \\
14 & 7200 & 10 & 0 & 3 & 0 & 240 & 700 & 0.20 & 0.20 & $9.51 \times 10^{-5}$ & $3.48 \times 10^{-1}$ & $3.57 \times 10^{8}$ & $3.23 \times 10^{7}$ & 64 \\
15 & 7200 & 10 & 0 & 4 & 0 & 1680 & 1980 & 0.14 & 0.14 & $1.61 \times 10^{-15}$ & $5.45 \times 10^{-8}$ & $2.44 \times 10^{8}$ & $2.41 \times 10^{7}$ & 82 \\
16 & 7200 & 10 & 0 & 1 & 0 & 1880 & 1480 & 0.18 & 0.18 & $1.43 \times 10^{-14}$ & $2.01 \times 10^{-4}$ & $5.61 \times 10^{8}$ & $9.88 \times 10^{7}$ & 77 \\
17 & 7200 & 10 & 0 & 2 & 0 & 2740 & 1040 & 0.15 & 0.15 & $1.96 \times 10^{-14}$ & $7.63 \times 10^{-5}$ & $3.74 \times 10^{8}$ & $1.70 \times 10^{7}$ & 72 \\
18 & 7200 & 10 & 0 & 1 & 0 & 1460 & 1720 & 0.15 & 0.15 & $1.96 \times 10^{-9}$ & $1.13 \times 10^{-3}$ & $2.10 \times 10^{8}$ & $2.56 \times 10^{7}$ & 78 \\
19 & 7200 & 10 & 0 & 5 & 0 & 1600 & 1720 & 0.14 & 0.14 & $6.55 \times 10^{-14}$ & $1.29 \times 10^{-4}$ & $2.67 \times 10^{8}$ & $4.09 \times 10^{7}$ & 75 \\
20 & 7200 & 10 & 0 & 3 & 0 & 4400 & 1280 & 0.19 & 0.19 & $3.27 \times 10^{-5}$ & $2.86 \times 10^{-2}$ & $3.26 \times 10^{8}$ & $2.27 \times 10^{7}$ & 64 \\
21 & 6000 & 0 & 50 & 92 & 88 & 2800 & 1000 & 2.64 & 0.13 & $1.97 \times 10^{-9}$ & $2.82 \times 10^{-3}$ & $1.67 \times 10^{8}$ & $5.23 \times 10^{7}$ & 45 \\
22 & 6000 & 0 & 50 & 79 & 91 & 1200 & 1400 & 4.56 & 0.17 & $2.72 \times 10^{-6}$ & $8.30 \times 10^{-3}$ & $7.56 \times 10^{7}$ & $2.80 \times 10^{7}$ & 53 \\
23 & 7200 & 10 & 50 & 90 & 100 & 840 & 1180 & 7.20 & - & $2.31 \times 10^{-5}$ & $4.17 \times 10^{-2}$ & $2.08 \times 10^{8}$ & $5.04 \times 10^{7}$ & 61 \\
24 & 7200 & 10 & 50 & 94 & 95 & 1080 & 1660 & 4.61 & 0.08 & $4.09 \times 10^{-4}$ & $6.70 \times 10^{-2}$ & $1.42 \times 10^{8}$ & $8.46 \times 10^{7}$ & 52 \\
\hline
\end{tabular}}
\end{table*}

\subsection{Remaining Lifetime: Time to Dissolution}
\label{time_dis_sect}

\begin{figure}
	\includegraphics[width=\columnwidth]{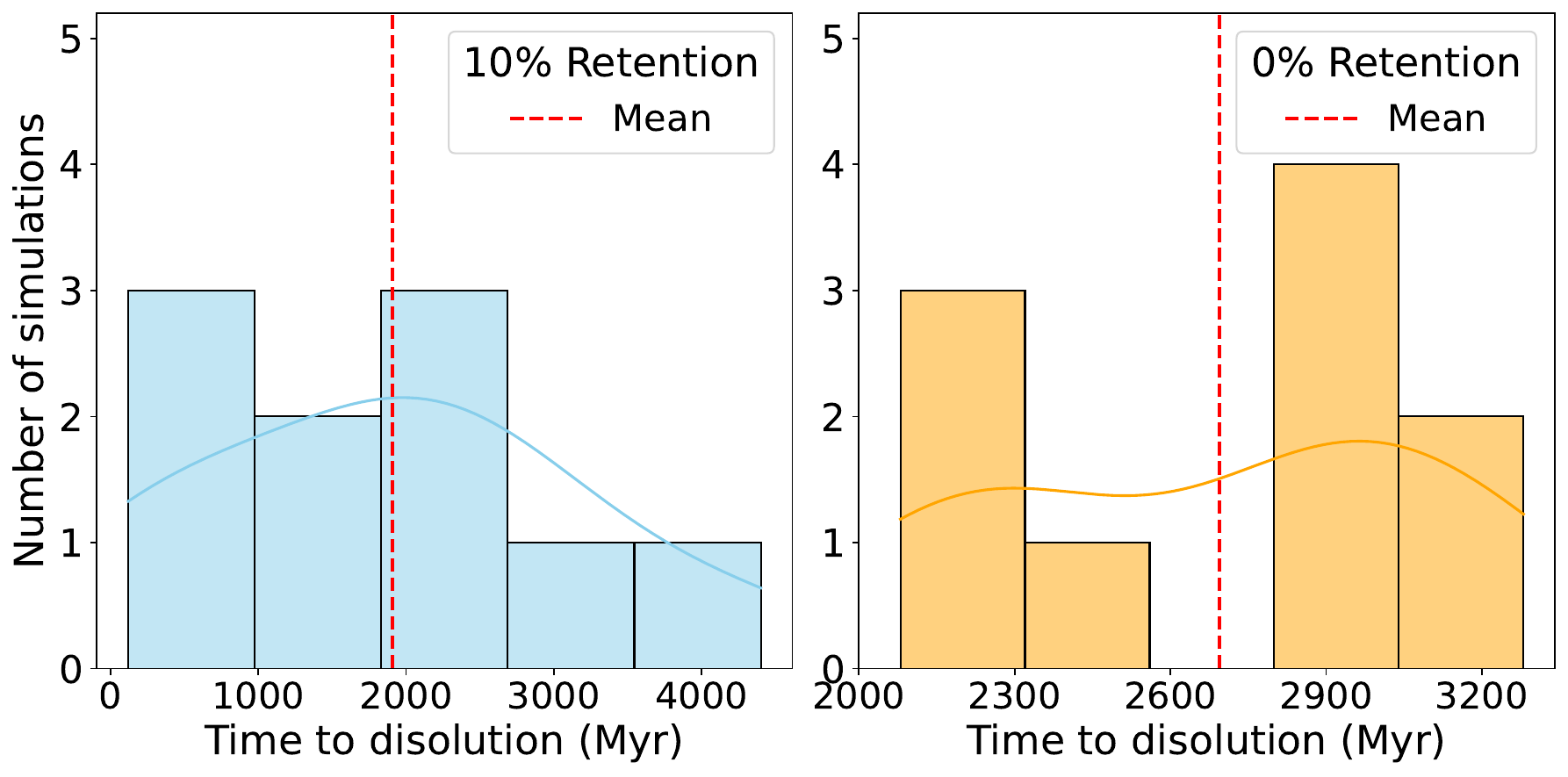}
    \caption{Left hand side: Distribution of the remaining lifetimes of the ten simulations with a 10\% black hole retention fraction and no primordial binaries.  Right hand side: The same for the ten simulations with a 0\% black hole retention fraction and no primordial binaries. The blue and yellow lines, are kernel density estimate trend lines. The 10\% retention simulations have a greater variability in their remaining lifetimes, due to increased stochasticity in the loss of heavy compact remnant binary stars.}
    \label{fig:time_to_diss}
\end{figure}

The remaining cluster lifetimes, $T_{diss,U}$ for our 20 $f_{b,0}=0$ simulations can be found in Table \ref{22_sims} and the distribution of these lifetimes is shown in Figure~\ref{fig:time_to_diss}. The remaining lifetime distributions of clusters 1 to 10 and clusters 11 to 20 satisfy the normality test outlined by \citet{shapiro_1965}, so we report only the mean with standard deviation. The average dissolution time for the ten simulations (11 to 20) with no primordial binaries and a 10\% retention rate of black holes and neutron stars is $<{T_{diss,U}}> = 1906 Myr \pm 1348 Myr$ and the average dissolution time for the ten simulations (1 to 10) with no primordial binaries and a 0\% retention rate is $<{T_{diss,U}}> = 2694 Myr \pm 432 Myr$. The 0\% retention simulations with no primordial binaries exhibit less variability in their dissolution times compared to the 10\% retention simulations with no primordial binaries. This can be attributed to the strong mass loss of the 10\% retention clusters once massive black hole binaries are lost from the clusters, as well as the variability in the black hole binary masses and lifetimes. An example of the 10\% retention simulations variability can be seen by comparing the two simulations depicted in Figure \ref{fig:time_evo}: The top two panels show the evolution of the bound mass in Simulation 12 from Table \ref{22_sims}. This simulation contains a massive black hole binary that is ejected from the cluster at 11740 Myr, causing a significant drop in the cluster's total mass and binding energy. As a result, the number of bound stars decreases from 42 to 13 within 20 Myr, and the cluster survives only for another 80 Myr. In contrast, simulation 17 (two bottom panels of Figure \ref{22_sims}), despite experiencing similar rapid drops in mass at earlier times (when the overall mass is larger), has a remaining single black hole binary. This remaining black hole binary allows the cluster to enter a more regular tidal stripping regime by the time it reaches UNIONS size (21 stars), resulting in a slower decay and a significantly longer time to dissolution.

All simulations contain many more stars than the UNIONS survey is able to detect, first main sequence stars outside of the UNIONS range, 0.31$M_{\odot}$< $M_u$ < 0.79 $M_{\odot}$, and secondly compact stellar remnants, especially white dwarfs. Once the clusters reach UNIONS size, compact remnants dominate the difference between the green (visible UNIONS range stars) and blue (all cluster stars) lines in Figure \ref{fig:time_evo}. The fraction of compact remnants in the cluster simulations at UNIONs time is found in column $f_{cr}$ in table \ref{22_sims}. The mean number of compact remnants in our 20 simulations without primordial binaries is $\overline{f_{cr}} = 74 \pm 1\%$.

Compact stellar remnants, predominantly white dwarfs formed through stellar evolution, comprise a substantial fraction of the mass of U1 and, through mass segregation, concentrate in the core of the cluster. This enhanced central mass concentration deepens the gravitational potential, increases the binding energy, and prolongs the cluster’s relaxation time, collectively extending U1’s remaining lifetime. In contrast, \citet{Errani_2024} did not include stellar evolution (and therefore omitted compact remnants) in their models, which largely explains why they predict a shorter remaining lifetime for UMa3/U1. Since the clusters are strongly dominated by compact remnants in the final stages, it would be very hard for an observer to detect a cluster just before its final dissolution, as there are very few bright stars left in the cluster at this point. We therefore calculate an alternative measure of the cluster's remaining lifetime (not considered by \citet{Errani_2024}: We estimate the duration for which UMa3/U1 would be the smallest cluster in the Milky Way's halo up to the point when UMa3/U1 becomes too faint for observers to see. We define this as the time required for UMa3/U1 to go from 50 bound stars within the UNIONS range to 10 bound stars in the UNIONS range, $T_{diss,U,50:10}$. This number range is chosen because a cluster with around 10 stars or less would not be visible as a star cluster, while a cluster with significantly more than 50 visible stars would not be considered exceptional since clusters with a slightly larger number of stars, such as Segue 3, are already known \citep{Belokurov_2010}. For the ten simulations with a retention fraction of 10\% and no primordial binaries ($f_{b,0}=0$), we find that the mean time for U1 to decrease from fifty to ten stars is \( <{T_{diss,U,50:10}}> = 1314 \, \text{Myr} \pm 548 \, \text{Myr} \), while for the ten simulations without primordial binaries and a retention fraction of 0\%, the mean time is \( <{T_{diss,U,50:10}}> = 1746 \, \text{Myr} \pm 356 \, \text{Myr} \) (uncertainties denote the standard deviation across simulations). Hence, for about 1 to 2 Gyr, U1 would be visible and smaller than any other known star cluster in the halo of the Milky Way. This duration represents a significant fraction of the cluster's total age and again argues against the idea that observing a system like U1 today would be highly unlikely if it were a star cluster. Varying the threshold for what constitutes a significantly small cluster to 40 stars instead of 50 has only a minor impact on the dissolution time, with both the 0\% and 10\% retention simulations still yielding average remaining lifetimes exceeding 1 Gyr.

\subsection{UMa3/U1 Density}
\label{density}

In addition to finding a short remaining lifetime of U1, \citet{Errani_2024} also conclude that their analytical mean density of U1, 
\(\rho_{u,Errani} = 2.9 \times 10^{7} \, M_{\odot} \, \text{kpc}^{-3}\) \citep[][equation 5]{Errani_2024} suggests that as a star cluster U1 is subject to rapid dissolution. This is because $\rho_{u,\mathrm{Errani}}$ is nearly equal to the mean density of the Milky Way at the orbital pericentre of UMa3/U1, located $12.8^{+0.8}_{-0.7}\,\mathrm{kpc}$ from the Galactic centre, which \citet{Errani_2024} estimated as $\rho_{\mathrm{Galaxy}} = 1.9 \times 10^{7}\,M_{\odot}\,\mathrm{kpc}^{-3}$.

However, the simulation code used in the \citet{Errani_2024} study, \textsc{Superbox}, does not account for individual stellar interactions. As a collisionless code, it cannot dynamically reproduce processes like mass segregation, which depend on two-body encounters and energy equipartition. These mechanisms, critical for the accumulation of massive compact stellar remnants in the core of a cluster, significantly enhance the central density. Only collisional codes like \textsc{Nbody7} model these effects correctly and can capture the contributions of compact stellar remnants to the cluster's evolution.

Table~\ref{22_sims} shows that the density derived solely from visible stars, \(\rho_{\rm HST}\), agrees reasonably well with \(\rho_{u,\rm Errani}\). However, when the contribution from compact remnants is taken into account (see the \(\rho_{\rm all}\) column), the inferred density increases by approximately an order of magnitude. This indicates that the actual density of UMa3/U1 is substantially higher than the Milky Way’s average density at the UMa3 / U1 pericentre, thus challenging \citet{Errani_2024}'s assertion that the U1 star cluster would quickly dissolve. This again demonstrates the importance of accounting for invisible compact stellar remnants, their effect on collisional dynamics and mass segregation when modelling dynamically highly evolved clusters like UMa3/U1.

\subsection{Velocity dispersion}
\label{vel_dispers}

\begin{figure}
	\includegraphics[width=\columnwidth]{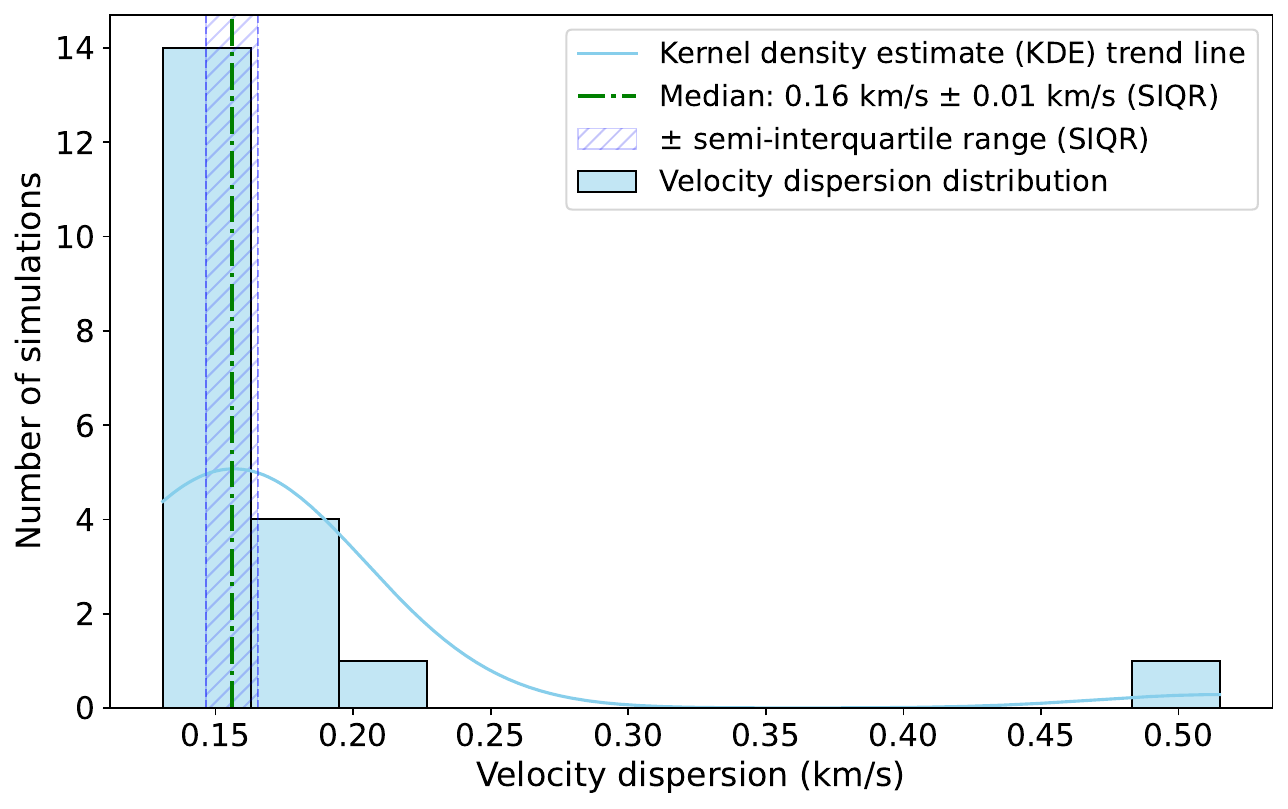}
    \caption{Distribution of velocity dispersion values of the clusters without primordial binaries, simulation 1 to 20, at UNIONS time in the UNIONS range. The smooth curves represent the kernel density estimates of the velocity dispersion distributions. To simulate observed velocity dispersions, individual velocity dispersion values are calculated using the luminosity-weighted average velocity of binary pairs in place of their individual velocities. (see section \ref{velocity dispersion method}). There is a tight distribution in velocity dispersion values around $0.16\,\mathrm{km\,s^{-1}}$, very close to the value expected for a star cluster similar in size and mass to U1 (Section \ref{vel_dispers}). Simulation 12 shows an outlying velocity dispersion of $0.51\,\mathrm{km\,s^{-1}}$, driven by a single star with an anomalously high velocity. This high velocity star is likely escaping the cluster at UNIONS time, having been disrupted from a binary system. }  
    \label{fig:velocity_dispersion}
\end{figure}

The mean velocity dispersion of the 20 $f_{b,0}=0$ star clusters at UNIONS size is, $\overline{\sigma}_{\mathrm{lum},U} = 0.18\,\mathrm{km\,s^{-1}}$ with a standard deviation of $\pm 0.08\,\mathrm{km\,s^{-1}}$. We calculate ${\sigma}_{\text{lum},U}$ using only the velocities of bright stars observable in the UNIONS survey, employing the luminosity-weighted average velocity method for binary star systems described by \citet[][method 3]{Rastello_2020}.  Figure~\ref{fig:velocity_dispersion} depicts the velocity dispersion distribution of the individual simulations, the values for which are found in Table \ref{22_sims}. Due to the non-normal data distribution, additional non-parametric measures are considered: a median velocity dispersion of 0.16$\,\mathrm{km\,s^{-1}}$ with an interquartile range of 0.02$\,\mathrm{km\,s^{-1}}$. We report the average velocity dispersion derived from simulations with both 0\% and 10\% retention fractions, as the exact retention fraction is uncertain: Previous studies indicate that clusters with escape velocities as low as UMa3/U1 ($4$ to $5\,\mathrm{km\,s^{-1}}$) should have low retention rates---possibly 0\%, and almost certainly below 10\% (e.g., \citealt{Morscher_2015}; \citealt{Hobbs_2005}). And though retained black holes and neutron stars significantly enhance the central velocity dispersion by deepening the gravitational potential of the core, the overall luminosity-weighted velocity dispersion remains largely unaffected \citep{Pavlik_2018}, thus analysing an average velocity dispersion from the combined set of simulations is valid.

Using the projected virial line of sight velocity dispersion formula from \citet{Errani_2018} and \citet{Errani_2024}:
\begin{equation}
\sigma_{\text{los}} = \sqrt{\frac{5}{96} \frac{GM_{\star}}{r_{\star}}} \,,
\label{analytic_dispersion}
\end{equation}
the average cluster mass, $M_{\star} = 142 \, M_{\odot}$ and the average half-mass radius from our simulations, $r_h = 4.8$ pc, with the 3D half-mass radius - scale radius relation; \( r_h \approx 2.67 \, r_{\star} \), an analytical velocity dispersion of $ 0.114\,\mathrm{km\,s^{-1}}$ is obtained, in good agreement with the velocity dispersion values of our simulations. We note that our estimated analytical velocity dispersion is over twice that of \citet{Errani_2024}, $\sigma_{los, Errani} = 0.049^{+0.014}_{-0.011}\,\mathrm{km\,s^{-1}}$, due to the significant fraction of the total cluster mass in compact remnants. However, both our simulated and analytical velocity dispersion values are significantly smaller than the observed velocity dispersion of UMa3/U1 as given by \citet{Smith_2024}, $\sigma_{\mathrm{los}} = 3.7^{+1.0}_{-1.4}\,\mathrm{km\,s^{-1}}$. 
Simulations 1 to 20 start without primordial binaries and develop on average only one binary pair per simulation by the time they reach UNIONS size. Furthermore, except for one simulation, none of these binaries contains a star bright enough to be visible to UNIONS. The average present-day binary fraction across all the UNIONS-visible stars in simulations 1 to 20, is \(\overline{f}_{\text{b,U}} = 0.47\%\), which is significantly lower than the binary fractions observed among field stars or what is expected in star clusters (see discussion in Section \ref{Introduction}). Thus, our first 20 simulations show that a significant primordial binary fraction is necessary to account for the likely high current-day binary fraction, consistent with the results of prior N-body studies \citep{Hutt_1992}, and that the low final binary count, \(\overline{f}_{\text{b,U}} = 0.47\%\) is insufficient to produce any discernible velocity dispersion inflation. Motivated by this unexpectedly low present-day binary fraction, we conduct and analyse four additional simulations (runs~21--24) incorporating a realistic primordial binary population, enabling a better evaluation of the effect of binaries on the cluster's velocity dispersion.

The average velocity dispersion for our four simulations with primordial binaries is $\overline{\sigma}_{\mathrm{lum,U}} = 4.75\,\mathrm{km\,s^{-1}}$. In comparison, the average velocity dispersion of single stars (excluding binaries) of the same simulations is $\overline{\sigma}_{\mathrm{sing,U}} = 0.12\,\mathrm{km\,s}^{-1}$. If the cluster's true dynamical mass is better represented by $\sigma_{\mathrm{sing,U}}$, then using the luminosity-weighted velocity dispersion, $\sigma_{\mathrm{lum,U}}$, would overestimate the cluster’s dynamical mass by a factor of approximately $1500$, given that the virial theorem implies the dynamical mass of an isotropic cluster in equilibrium scales as the square of its velocity dispersion. Such an overestimation is sufficient to reduce the observed dynamical mass-to-light ratio of $M/L = 6500\,M_{\odot}/L_{\odot}$ \citep{Smith_2024} to levels consistent with a dispersion supported stellar cluster without significant dark matter. 

All four primordial binary simulations showed a significant increase in binary fraction, and were composed of $79\%$ to $94\%$ binaries at UNIONS time. Although globular clusters tend to have their binary fraction decrease over time \citep{Heggie_1975}, the much less dense stellar environment of UMa3/U1 has fewer three-body encounters and collisions than a typical dense globular cluster; therefore, fewer binaries are disrupted. Thus, the effect of the relatively heavy binary and higher order systems sinking to the cluster core via energy equipartition and the preferential stripping of single (lighter) systems dominates over the binary disruption rate resulting in an increasing concentration of binaries in the core over time.

\begin{figure}
	\includegraphics[width=\columnwidth]{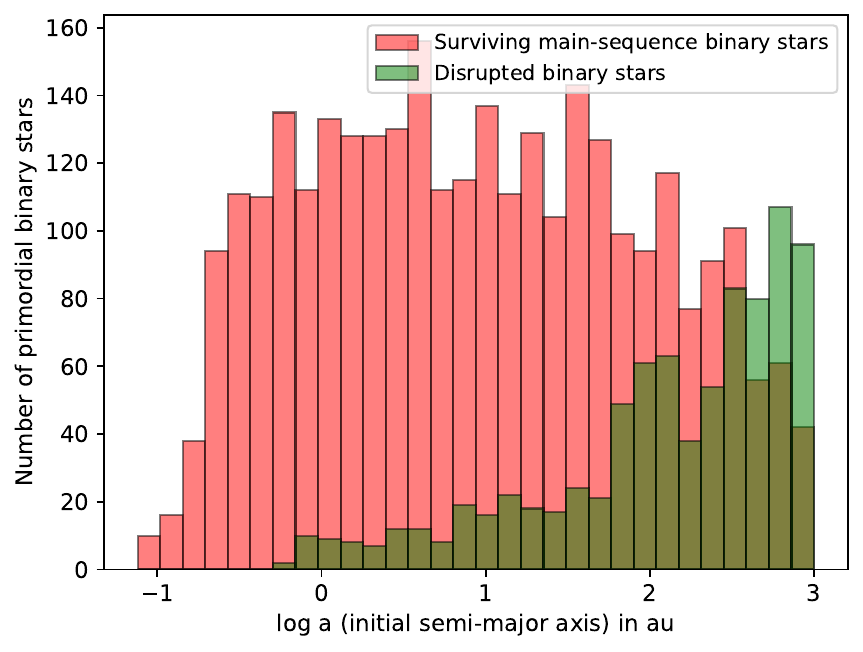}
    \caption{Distribution of initial semi-major axis of surviving primordial binaries (red) and primordial binaries that have been broken up by UNIONS time (green), for stars in simulation 23. The survival rate of binaries of different initial semi-major axis can be inferred by the difference between the red and green bars. For binaries with separations less than approximately 380~au, most binaries survive.}
    \label{fig:binary survive}
\end{figure}

The strong survival rate of hard binaries is illustrated in Figure \ref{fig:binary survive}, which shows the distribution of the initial semi-major axes of the primordial binaries within the cluster and its tails. The figure demonstrates that binaries with initial semi-major axes approximately less than 380 au are more likely to survive than to be disrupted. Although the focus here is on binary survival within the cluster, Figure \ref{fig:binary survive} includes both cluster members and stars located in the tidal tails, rather than cluster stars alone. Tidal tail stars are included because, by UNIONS time, the few remaining single stars within the cluster are not the result of disrupted primordial binaries. Furthermore, the majority of systems found in the tidal tails, whether single, binary, or higher-order, have spent a significant portion of their lifetimes within the cluster. Therefore, the distribution shown in Figure \ref{fig:binary survive} remains representative of the binary survival rate in the cluster environment.

\begin{figure*}
    \centering
    \begin{minipage}{0.48\textwidth}
        \centering
        \includegraphics[width=\columnwidth]{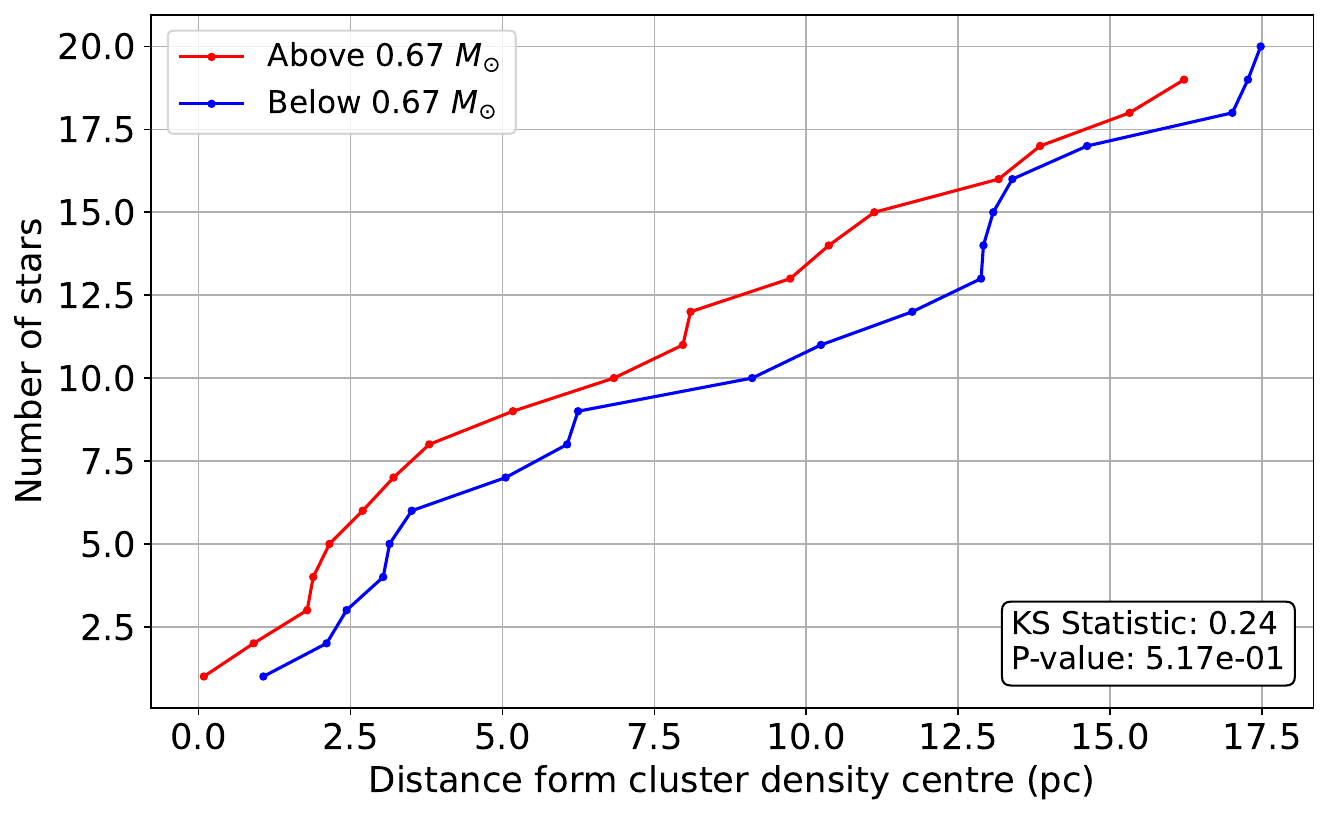}
    \end{minipage}
    \hfill
    \begin{minipage}{0.48\textwidth}
        \centering
        \includegraphics[width=\columnwidth]{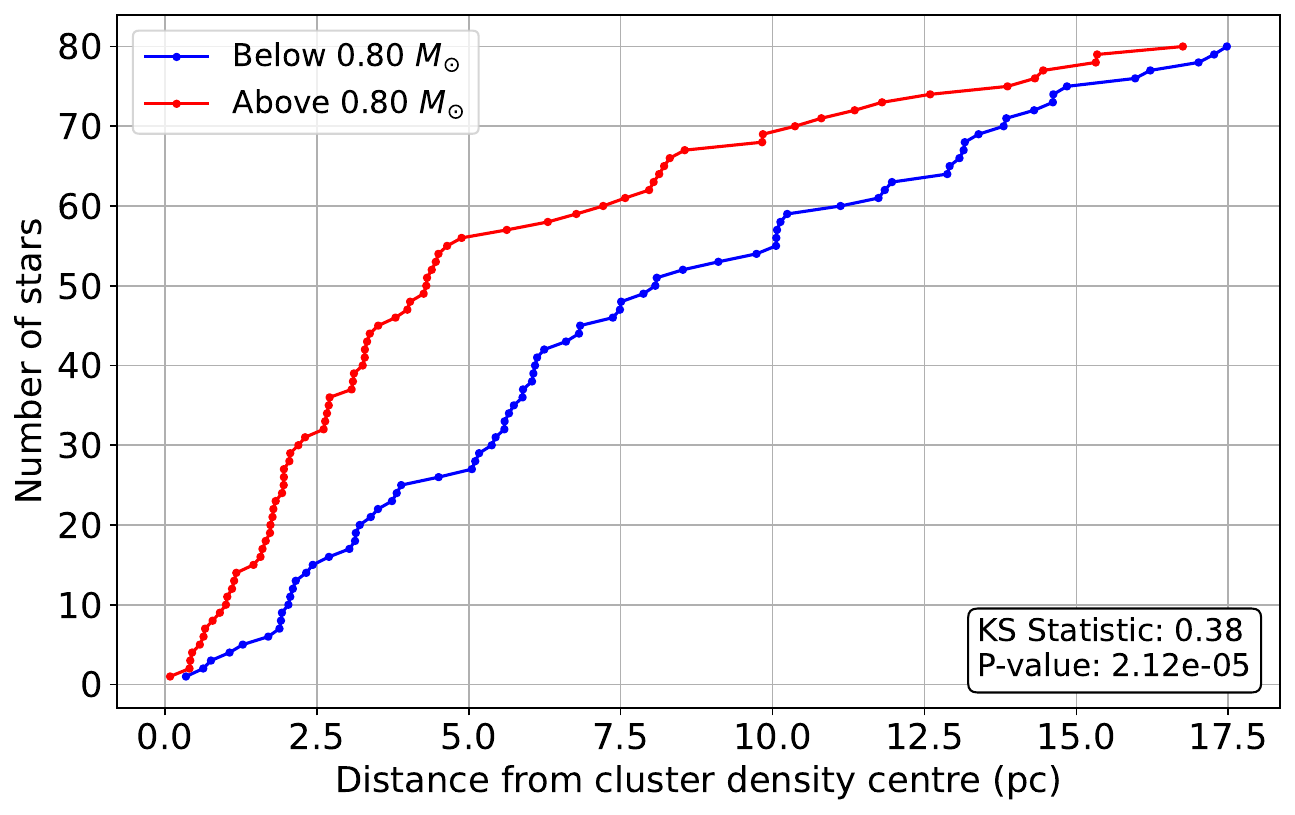}
    \end{minipage}
    \caption{Cumulative radial distribution of bound stars in simulation 3 at UNIONS time, T=12140 Myr. The left panel shows the distribution of \textit{HST} range stars while the right panel shows the distribution of all bound stars (including compact remnants). The lower curves show the cumulative distribution of stars with masses below the median, while the upper curves show the cumulative distribution of stars with masses above the median.  When restricted to \textit{HST} stars, the significance of the mass segregation is dramatically reduced because the cluster consists mostly of compact stellar remnants not included in the \textit{HST} range.}
    \label{fig:side_by_side}
\end{figure*}

The markedly increased velocity dispersion, $\sigma_{\text{lum,U}}$, observed in our primordial binary simulations (simulations 21 - 24) compared to $\sigma_{\text{sing,U}}$ across all simulations suggests that binary stars can significantly enhance the velocity dispersion of UMa3/U1. Consequently, the high velocity dispersion reported by \citet{Smith_2024} does not necessarily indicate that UMa3/U1 is a dark matter-dominated dwarf galaxy. This underscores the importance of multiepoch spectroscopy to derive a velocity dispersion that accurately reflects the true mass-to-light ratio of UMa3/U1.

\subsection{Mass Segregation} 
\label{mass_seg_section}

\citet{Baumgardt_22} conduct a comprehensive analysis of the degree of mass segregation in more than 50 globular clusters (GC) and ultrafaint dwarf galaxies (UFD), revealing that GCs with ages equal to or exceeding their relaxation times exhibit significant mass segregation. Given that most GCs are older than their relaxation times, mass segregation is a common characteristic among them. In contrast, UFDs generally have much longer relaxation times due to their large dark matter content and do not exhibit mass segregation. Consequently \citet{Baumgardt_22} conclude that mass segregation serves as a valuable distinguishing criterion between globular clusters and UFDs.

To determine whether mass segregation should be expected in our model cluster U1, we estimate the relaxation times according to \citet{Spitzer_1987}:
\begin{equation}
T_{\text{RH}} = 0.138 \frac{\sqrt{M} r_h^{1.5}}{\sqrt{G} \langle m \rangle \ln(\gamma N)},
\end{equation}
where $M$ is the (stellar) mass of the system, $r_h$ the three-dimensional half-mass radius, $\langle m \rangle$ the average mass of stars, and $N$ is the number of stars. $\gamma$ is a constant in the Coulomb logarithm for which we assume $\gamma = 0.11$ \citep{Giersz_1994}. We find mean relaxation times of
\(\overline{T}_{\text{RH}} = 90\) Myr and \(\overline{T}_{\text{RH}} = 158\) Myr for the 0\% and 10\% retention models, respectively. Therefore, if UMa3/U1 is a cluster, it is very likely mass segregated. However, proving mass segregation using known stars is challenging because of the limited mass range and the small number of UNIONS observable stars.

To mitigate the problem of small number statistics and in order to include stars of lesser mass to increase the mass range, we assume that UMa3/U1 would be the target for a space telescope like the \textit{HST} or the \textit{JWST}, which would allow observation of main sequence stars nearly down to the hydrogen burning limit. We therefore analyse our simulations in the \textit{HST} filter which takes into account all main sequence and giant stars with masses as low as 0.1 $M_\odot$ (the lowest mass used in our simulations). In Figure~\ref{fig:side_by_side} we plot the cumulative radial distribution of stars below and above the median mass for simulation 3. It is clear from the cumulative distribution plots that mass segregation exists in the cluster (as expected), but it is less pronounced when observations are limited to the \textit{HST} range. Furthermore, most simulations within the \textit{HST} range do not show clear mass segregation. To quantify the mass segregation, a Kolmogorov-Smirnov (KS) test is performed, comparing the cumulative distributions of low and high mass stars to assess the significance of the differences between these distributions \citep{Press_1992}. The average KS test statistic and the $p$-value for our $f_{b,0}=0$ simulations are found in Table~\ref{tab:mass_segregation_statistics}. When including all bound stars, 15 out of our 20 $f_{b,0} = 0$ simulations show statistically significant ($p$-values lower than the significance level of 0.05) mass segregation. However, when compact remnant stars are excluded and only stars within the \textit{HST} detection range are considered, mass segregation becomes notably less significant—only 2 out of 20 simulations show mass segregation with \(p < 0.05\). This reduction in significance arises because compact remnant stars constitute the majority of the cluster population at UNIONS time, so their inclusion enhances the statistical robustness of the mass segregation signal.

Based on these findings, we conclude that if UMa3/U1 is a self-gravitativing star cluster, it is likely to be mass segregated, as its relaxation time is much shorter than its age, and our N-body simulations show some mass segregation. But mass segregation is not currently a useful classification tool for UMa3/U1 because our N-body simulations show that the subset of stars seen by deep-photometry surveys (i.e. main sequence and giant stars) do not exhibit mass segregation at statistically significant levels.

\begin{table}
	\centering
	\caption{Average mass segregation statistics for simulations 1–10 (0\% retention) and simulations 11–20 (10\% retention). Results are shown first for all bound stars, and then for bound stars excluding compact remnants (\textit{HST} range stars). Uncertainties represent the standard deviation across simulations. The median and semi-interquartile range (SIQR) are included to account for potential non-normality in the distributions.}
	\label{tab:mass_segregation_statistics}

\begin{tabular}{@{}lcccc} 
\hline
RR & Metric & Average Value & Median & SIQR\\
\hline
0\% (all bound stars) & p-value & $0.0091 \pm 0.0250$ & 0.0002 & 0.0009\\
                      & KS-stat & $0.3138 \pm 0.0527$ & 0.3246 & 0.0190\\
10\% (all bound stars) & p-value & $0.1566 \pm 0.3125$ & 0.0270 & 0.0432\\
                      & KS-stat & $0.2426 \pm 0.0777$ & 0.2582 & 0.0564\\
0\% (\textit{HST} bound stars)   & p-value & $0.4661 \pm 0.2792$ & 0.4974 & 0.2232\\
                      & KS-stat & $0.2839 \pm 0.0666$ & 0.2513 & 0.0479\\
10\% (\textit{HST} bound stars)   & p-value & $0.3167 \pm 0.3111$ & 0.1719 & 0.2312\\
                      & KS-stat & $0.3340 \pm 0.1230$ & 0.3419 & 0.0418\\
\hline
\end{tabular}
\end{table}

\subsection{Mass function}
\label{mass_funct_sect}

Although several studies estimate the effects of tidal stripping of stars and dark matter from UFDs \citep{Penarrubia_2008, Errani_2018, Fattahi_2016}, it is well documented that UFDs are among the most dark matter dominated satellites of the Milky Way, making them highly resistant to tidal stripping \citep{Belokurov2007, Simon_2007, Willman_2010, Errani_2015, Simon_2019}. Given that UMa3/U1 is the smallest known UFD candidate, and since the size and luminosity of UFDs are inversely correlated with their mass-to-light ratios \citep{Simon_2007, Martin_2008, Wolfe_2010}, 
it is likely that if UMa3/U1 is a UFD, it will have experienced little mass loss and, therefore, will display little to no change in its stellar mass function. Even with some mass loss, its mass function has likely not changed because dark-matter-dominated systems like UFD's are resilient to mass segregation, and thus the stars on the outside of the cluster, those most likely to be stripped first, are well mixed in mass compared to a mass segregated system. Consequently, UMa3's present-day mass function is expected to closely resemble its initial mass function. However, star clusters, owing to their much lower relaxation times, are much more likely to be mass segregated and therefore to have experienced preferential loss of low-mass stars by the external tidal field, resulting in their mass function becoming top heavy over time \citep{Baumgardt_2003}. We therefore evaluate the effectiveness of the stellar mass function as a criterion for differentiating UFDs from star clusters by comparing the mass function of the cluster simulations when they have reached UNIONS time against a modelled UMa3 UFD mass function. We evaluate mass functions in both the UNIONS and \textit{HST} range.  

To model the UMa3 UFD, we sample from the same initial mass function which we use to create our progenitor U1 clusters (Section \ref{progen_model}). We compare the UMa3 model mass function with each star cluster’s mass function using the non-parametric Kolmogorov–Smirnov (KS) test; the resulting statistical p-values are listed in Table~\ref{22_sims}. A significant difference (\textit{p} value less than 0.05) is seen in 20 of the 20 $f_{b,0}=0$ simulations, when using stars within the \textit{HST} range, but only 17 of the 20 $f_{b,0}=0$ simulations when using stars within the UNIONS range. So, for the mass function test to be consistently significant, deeper photometry is necessary. 

To determine the photometric depth required for the mass function test to become statistically significant, we conduct multiple mass function tests, altering the mass range on each test, to include stars down to the mass equivalent of a range of faint magnitudes. We convert masses to magnitudes in the F814W filter of the Ultraviolet--Visible (UVIS) channel of the \textit{HST} Wide Field Camera~3 (WFC3) using the PARSEC~1.2 isochrones \citep{Bressan_2012}, adopting parameters identical to those listed in items~(i)--(iii) of Section~\ref{sec:Unions_range}. We find that achieving an $i$-band magnitude of 25 in the F814W filter enables a robust mass function test: at this magnitude, corresponding to approximately $0.23\,M_\odot$, all simulations yield a KS-test \textit{p}-value substantially below 0.05. Calculations performed using the Space Telescope Science Institute's (STScI) WFC3/UVIS Imaging Exposure Time Calculator \citep{WFC3UVIS_ETC} indicate that reaching a depth of $i = 25$\,mag requires less than 1~hr of observation time, assuming observations are conducted in two filters.

In conclusion, the mass function test as outlined in this section used with photometry of $i = 25$\,mag or deeper will be able to classify UMa3/U1 as either a UFD or self-gravitating star cluster. Furthermore, it can be used to classify other small, faint Milky Way satellites. We caution that the mass function classification method may be less effective in classifying dynamically young or distant clusters, since such systems experience less tidal stripping and therefore exhibit smaller deviations in their present-day mass functions relative to their initial mass functions (IMFs). Consequently to establish meaningful limits on the applicability of this classification tool, we recommend conducting simulations of star clusters that have undergone less mass loss than U1.

\begin{figure}
	\includegraphics[width=\columnwidth]{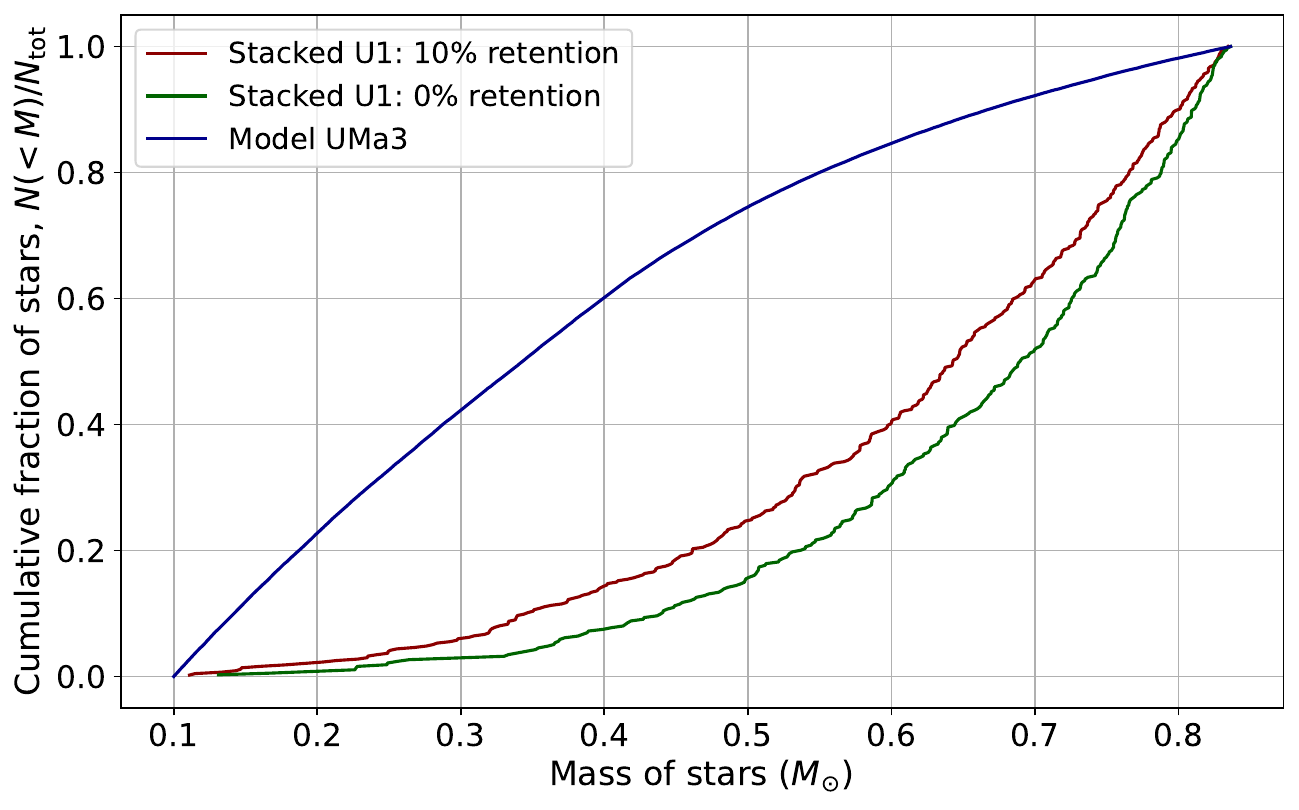}
    \caption{Cumulative mass function of \textit{HST} range stars. The lowermost curve shows the stacked simulations with 10\% black hole and neutron star retention and no primordial binaries. The middle curve shows the stacked simulations with 0\% retention and no primordial binaries. The top curve represents the UMa3 ultra-faint dwarf (UFD) galaxy model. There is a significant difference between the mass function of UMa3/U1 modelled as a dwarf galaxy and that of the simulated star clusters}
    \label{fig:m_funct}
\end{figure}

To demonstrate the statistical robustness of the mass function result we also stack the present day \textit{HST} range mass function of simulations 1 to 10 (0\% retention rate) and simulations 11 to 20 (10\% retention rate) to obtain better statistics, again comparing the stacked cluster simulations to an equivalent sized UMa3 UFD model mass function, modelled from a \citet{Baumgardt_2023} IMF. The cumulative mass function plot of the stacked simulations, Figure \ref{fig:m_funct}, clearly shows a significant difference in the mass functions of the UMa3 UFD model and the stacked U1 models. The stacked cluster mass functions appear significantly more top-heavy than the UMa3 UFD model mass function, reflecting a concentration of massive, predominantly compact remnant stars. Comparing the stacked 10\% retention models without primordial binaries (simulations 11 to 20) to the UFD model, the KS statistic is 0.51 and the $p$-value is $1.51 \times 10^{-102}$. Comparing the stacked 0\% retention cluster mass functions with the UFD mass function model, the KS statistic is 0.60 and the $p$-value is $ 1.88 \times 10^{-126}$. Taking the least significant of the two results gives a significance level of approximately $\sigma = 21.5$. 

We give numerical criteria to the expected present day mass function of U1, by fitting a single power law. We use the Maximum Likelihood method (MLE) following \citet{Khalaj_2013} to obtain the most likely slope of the present day mass functions, $\alpha$:
\begin{equation}
\alpha = 1 + n \left[ \left( \sum_{i=1}^{n} \ln\left(\frac{x_i}{x_{\min}}\right) \right) - n \cdot \frac{\ln X}{1 - X^{\alpha - 1}} \right]^{-1}
\end{equation}
as well as the error:
\begin{equation}
\sigma(\alpha) = \frac{1}{\sqrt{n}} \left[ (\alpha - 1)^{-2} - \ln^2 X \cdot \frac{X^{\alpha - 1}}{(1 - X^{\alpha - 1})^2} \right]^{-1/2}
\end{equation}

Here, \( x_i \) denotes the mass of the \( i \)-th star in the sample, \( x_{\min} \) and \( x_{\max} \) are the lower and upper mass limits of the fitting range, respectively, \( X = x_{\max}/x_{\min} \), and \( n \) is the number of stars in the sample. We find an alpha value of 2.53 for the stacked 0\% retention simulation and 1.46 for the stacked 10 \% simulations. So if UMa3/U1 is a star cluster we expect its mass function to fit a positive single power law slope similar to slopes depicted by Equations \ref{0.0 retention PDMF} and \ref{0.1 retention PDMF}: 
\begin{equation}
\label{0.0 retention PDMF}
\xi(m) \, dm \sim m^{2.53 \pm 0.19} \, dm, \quad \text{for } 0.13 \, M_{\odot} \leq m < 0.84 \, M_{\odot}
\end{equation}
\begin{equation}
\label{0.1 retention PDMF}
\xi(m) \, dm \sim m^{1.46 \pm 0.13} \, dm, \quad \text{for } 0.11 \, M_{\odot} \leq m < 0.86 \, M_{\odot}
\end{equation}
But if UMa3/U1 is a UFD, it should be better fitted by a negative slope, consistent with the IMF used to model UMa3. (equation \ref{IMF}).

\section{Discussion \& Conclusion}
\label{summarry_sect}

In this paper, we conduct dynamical N-body simulations of UMa3/U1 as if it were a star cluster to determine its remaining lifetime and, by extension, the validity of the recent claim that a short modelled remaining lifetime implies that UMa3/U1 is a UFD. We perform further analysis to determine the techniques and observations required for the successful classification of UMa3/U1.

We find a high velocity dispersion for clusters with primordial binaries, $\sigma_{\mathrm{lum,U}}=4.75\,\mathrm{km\,s}^{-1}$, indicating that the velocity dispersion measured by \citet{Smith_2024} for UMa3/U1, \( \sigma_{los}=3.7^{+1.0}_{-1.4} \, \text{km} \, \text{s}^{-1} \), is not inconsistent with a star cluster classification. Further, we find that the average remaining lifetimes of U1 are between 1.9 Gyr and 2.7 Gyr depending on the assumed retention rate of black holes and neutron stars against natal kicks. These remaining lifetimes are significantly longer than the 0.4 to 0.8\,Gyr remaining lifetime calculated by \citet{Errani_2024}, and this suggests that UMa3/U1 may continue to survive for a significant fraction of its proposed age, opening up the possibility that UMa3/U1 is a star cluster. The remaining lifetime found by \citet{Errani_2024} is shorter, primarily due to the significant underestimation of the total cluster mass by \citet{Smith_2024}. The mass underestimate arises because compact remnants make up roughly 50 to 80\% of cluster mass (Table \ref{22_sims}), yet these stars are invisible to the UNIONS survey. Our analysis accurately incorporates this high fraction of compact remnants concentrated at the cluster's centre by utilising a collisional N-body program that includes stellar evolution and a realistic spread of stellar masses. By properly accounting for close encounters, two-body relaxation is correctly modelled, allowing for an accurate representation of energy equipartition. The resulting mass segregation causes heavier stars, predominantly compact remnants, to concentrate toward the cluster centre. Consequently, our modelled clusters possess greater binding energy, resulting in significantly longer remaining lifetimes.

We choose to highlight the time to dissolution, $T_{diss,U}$ of the 0\% retention simulations without primordial binaries (simulations 1 to 10) in our abstract for several reasons: Although supernova theory suggests that some natal kick mechanisms might be more symmetric at certain masses (resulting in velocities below the UMa3/U1 escape velocity of approximately 4 to 5$\,\mathrm{km\,s^{-1}}$ \citep{Fryer_2012, Banerjee_2020}), observations of neutron stars and black holes with such low velocities have not been made. Whilst the lack of observed low velocity black holes and neutron stars could be due to selection effects, studies like \citet{Hobbs_2005} indicate that the frequency of such low velocities is expected to be less than 1\%. Additionally, slight selection effects may have been introduced in the 10\% retention rate simulations without primordial binaries (simulations 11 to 20) because the average total remaining lifetime for these simulations is 11.7 Gyr, slightly below our target age of 12 Gyr, hence the accurate initial size of the 10\% retention clusters is likely around 7350 stars.

In \citet[][appendix A]{Errani_2024} alternative Milky Way potential models are explored. Here Errani et al. also account for observational uncertainties in the present position and velocity of UMa3/U1. Different remaining lifetime estimates are achieved from those in their abstract and summary; for example, by using a \citet{Bovy_2015} potential, Errani et al. obtain a survival time of approximately 1.4 Gyr. This shows that significant variability in remaining lifetime can also be obtained by altering the cluster orbit and Milky Way potential, which warrants further investigation.

Our N-body simulations also show that if UMa3/U1 is a star cluster, it would likely exhibit mass segregation. However, it is unlikely that UNIONS or even deeper photometric surveys will be able to observe significant mass segregation in UMa3/U1 because of their inability to detect faint compact remnant stars at the 10~kpc distance of UMa3/U1.
Instead, the present-day mass function test provides a robust method for classifying UMa3/U1 as either an ultra-faint dwarf galaxy (UFD) or a self-gravitating star cluster. Consequently, we recommend that future observational studies of UMa3/U1 obtain photometric data down to an apparent magnitude of 25 and apply the mass function test outlined in this paper. With ongoing and upcoming surveys such as the Legacy Survey of Space and Time (LSST) by the Vera C. Rubin Observatory, the Dark Energy Survey (DES), and Pan-STARRS expected to uncover numerous faint Milky Way satellites over the next decade, the present-day mass function holds significant potential for revealing the true nature of these systems and thereby advancing cosmological models.

\section*{Acknowledgements}

SMS acknowledges funding from the Australian Research Council (DE220100003).
Parts of this research were conducted by the Australian Research Council Centre of Excellence for All Sky Astrophysics in 3 Dimensions (ASTRO 3D), through project number CE170100013.

\section*{Data Availability}

The data used in this study are available on request by emailing the first author.



\bibliographystyle{mnras}
\bibliography{example} 





\bsp	
\label{lastpage}
\end{document}